\documentclass[usenatbib,usegraphicx]{mn2e}
\usepackage{amsmath}
\usepackage{bm}
\usepackage{caption}
\usepackage{hyperref}
\usepackage[pdftex]{color}
\usepackage{subfigure}
\usepackage{xcolor}

\hypersetup{
    colorlinks,
    linkcolor={red},
    citecolor={blue}, 
    urlcolor={purple}
}

\title[The open cluster Praesepe]
{The stellar mass function, binary content and radial structure of the open cluster Praesepe derived from PPMXL and SDSS data}

\author[P. Khalaj and H. Baumgardt]{P. Khalaj\thanks{E-mail:
pouria.khalaj@uqconnect.edu.au} and H. Baumgardt\\
School of Mathematics and Physics, University of Queensland, St. Lucia, QLD 4072, Australia}

\begin{document}

\date{Accepted 2013 July 05. Received 2013 July 05; in original form 2013 February 24}

\pagerange{\pageref{firstpage}--\pageref{lastpage}} \pubyear{2013}

\maketitle

\label{firstpage}

\begin{abstract}
We have determined possible cluster members of the nearby open cluster Praesepe (M44) 
based on $J$ and $K$ photometry and proper motions from the PPMXL catalogue and $z$ photometry from the Sloan Digital Sky Survey (SDSS).
In total we identified 893 possible cluster members down to a magnitude of $J=15.5$\ mag, corresponding to 
a mass of about 0.15\,M$_\odot$ for an assumed cluster distance modulus of $(m-M)_0=6.30$\,mag ($d\approx182$\,pc), within a radius 
of 3.5$^\circ$ around the cluster centre. We derive a new cluster centre for Praesepe ($\alpha_{\rm centre}=8^h 39^m 37^s , \delta_{\rm centre} = 19^\circ 35' 02''$). We also derive a total cluster mass of about 630\,M$_\odot$ and a 2D half-number and half-mass radius of $4.25$\,pc and $3.90$\,pc respectively. The global mass function (MF) of the cluster members 
shows evidence for a turnover around $m=0.65$\,M$_\odot$.  While more massive stars can be fit by a power-law $\xi(m) \sim m^{-\alpha}$ with 
slope $\alpha=2.88\pm0.22$, stars less massive than $m=0.65$ M$_\odot$ are best fitted with 
$\alpha=0.85\pm0.10$. In agreement with its large dynamical age, we find that Praesepe is strongly
mass segregated and that the mass function slope for high mass stars steepens from a value of $\alpha=2.32\pm0.24$ 
inside the half-mass radius to $\alpha=4.90\pm0.51$ outside the half-mass radius.
We finally identify a significant population of binaries and triples in the colour-magnitude 
diagram of Praesepe. Assuming non-random pairing of the binary components, 
a binary fraction of about 35\% for primaries in the mass range $0.6 < m/\rm M_\odot < 2.20$
is required to explain the observed number of binaries in the colour-magnitude diagram (CMD). 
\end{abstract}


\section{Introduction} \label{sec:introduction}

Open clusters are important test beds for star formation, stellar initial mass function (IMF) and stellar evolution theories \citep{Gaburov}, since they 
provide statistically significant samples of stars of known distance, age and metallicity. The identification
of cluster members is especially easy in nearby clusters ($d<200$\,pc) since these have on average large proper motions
which allow to effectively separate cluster members from background stars. With a distance modulus of $(m-M)_0=6.30\pm0.07$\,mag 
\citep{van Leeuwen} ($d=181.97_{-5.77}^{+5.96}$\,pc) Praesepe is one of the nearest clusters to the Sun.  
Due to its proximity, Praesepe has been studied extensively in the past. However, so far
no consensus on the stellar distribution and low-mass mass function has been reached. Proper motion
studies of the bright members were first carried out by \cite{Klein}, \cite{Jones83} and \cite{Jones91}, who determined members down to $V \sim 18$\,mag,
corresponding to masses of about 0.3 M$_\odot$. \cite{Hambly} presented cluster members down to 0.1 M$_\odot$ in the central 19\,deg$^2$ using images taken by the United Kingdom Schmidt Telescope (UKST). They found evidence for mass segregation in Praesepe and a rising mass function from 1\,M$_\odot$ down to 0.1\,M$_\odot$, the limit of their survey. \cite{Pinfield} conducted a photometric survey of Praesepe in $RIZ$-bands down to $I=21.5$ using the Isaac Newton Telescope (INT). They also found a rising mass function ($\xi(m) \sim dN/dm \sim m^{-\alpha}$) from 0.15\,M$_\odot$ down to 0.07\,M$_\odot$ with slope $\alpha\geq1.5$. In contrast, using proper motions derived from the Two Micron All Sky Survey (2MASS) and the Palomar Observatory Sky Survey, \cite{Adams} found that the low-mass stellar mass function below 0.4\,M$_\odot$ can be fitted by a flat mass function with $\alpha \approx 0$ and only a marginal radial dependence of the mass function. \cite{Chappelle} probed the central 2.6\,deg$^2$ of Praesepe using $I$-band data down to $I\sim 21.3$\,mag and $Z$-band data down to $Z\sim 20.5$\,mag, from images taken by the INT/ Wide Field Camera (WFC) and near-infrared follow-up measurements using the United Kingdom Infrared Telescope (UKIRT) Fast Track Imager (UFTI). They found a rising mass function from 1\,M$_\odot$ down to 0.1\,M$_\odot$. \cite{Gonzalez} conducted deep photometric searches for sub-stellar members of Praesepe using the Sloan $i'$ and $z'$ broad-band filters, with the 3.5-meter and the 5-meter Hale telescopes on the Calar Alto and Palomar Observatories. The total area that they surveyed was 1177 arcmin$^2$ and the $5\sigma$ detection limit of their survey was $i'= 24.5$\,mag and $z' = 24$\,mag, which corresponds to $m=50-55$\,M$_{\rm Jup}$. They found that the mass function of Praesepe strongly depends on the adopted cluster age and at the youngest possible ages of Praesepe (500-700 Myr), their analysis suggests a rapidly decreasing mass function for brown dwarfs. \cite{Kraus} combined archival survey data from SDSS, 2MASS, USNO-B1.0 and UCAC-2.0 and found 1010 stars in Praesepe as candidate members with probability $p>80\%$. Their result for the mass function of Praesepe is similar to that of \citet{Hambly}, i.e a rise from 1\,M$_\odot$ down to 0.1\,M$_\odot$. \cite{Boudreault10} performed an optical ($I_{\rm c}$-band) and near-infrared (J and $K_{\rm s}$-band) photometric survey of the innermost 3.1\,deg$^2$ of Praesepe with $5\sigma$ detection limits of $I_{\rm c}=23.4$\,mag and $J=20.0$\,mag. They observed that the mass function of Praesepe rises from 0.6\,M$_\odot$ down to 0.1\,M$_\odot$ with $\alpha=1.8\pm0.1$ and turns over at $\sim 0.1$\,M$_\odot$. \cite{Baker} found a moderately rising mass function with $\alpha=1.11\pm0.37$ for the mass range $0.6$\,M$_\odot$ to $0.125$\,M$_\odot$ in the UKIRT Infrared Deep Sky Survey Galactic Clusters Survey (UKIDSS GCS) for $Z<18$. Using observations from the Large Binocular Telescope (LBT) in the $rizY$ bands, \cite{Wang} identified 62 cluster member candidates (40 of which are sub-stellar) within the central 0.59 deg$^2$ of Praesepe down to a $5\sigma$ detection limit of $i \sim 25.6$\,mag ($\sim 40$\,M$_{\rm Jup}$). They found that the mass function of Praesepe shows a rise from $105$\,M$_{\rm Jup}$ to $60$\,M$_{\rm Jup}$ and then a turnover at $\sim 60$\,M$_{\rm Jup}$. More recently, \cite{Boudreault12} found 1116 cluster candidates in a $\sim 36$\,deg$^2$ field based on a $3\sigma$ astrometric and five-band ($ZYJHK$) photometric selection, using the Data Release 9 (DR9) of UKIDSS GCS. They found that the mass function of Praesepe has a maximum at $\sim 0.6$\,M$_\odot$ and then decreases to the lowest mass bin of 0.056\,M$_\odot$.

In the present paper, we determine Praesepe members based on the PPMXL catalogue \citep{Roeser10} and SDSS DR9.
Our paper is structured as follows: we first discuss the observational data in section \ref{sec:obs}. In section 
\ref{sec:memdet} we discuss the procedures we have followed to determine possible members. Our results are presented in section \ref{sec:results} and 
we finally summarize our work in section \ref{sec:conclusion}.

\section{Observational Data} \label{sec:obs}
In this study we combine data from the PPMXL catalogue \citep{Roeser10} with $z$ magnitudes from SDSS DR9 \citep{Ahn}. 
PPMXL catalogue combines the USNO-B1.0 \citep{Monet} and 2MASS catalogues \citep{Skrutskie} yielding the largest collection of proper motions in the International Celestial Reference Frame (ICRS) to date \citep{Roeser10}. USNO-B1.0 contains the positions of more than one billion objects taken photographically around 1960; 2MASS is
an all-sky survey conducted in the years 1997 to 2001 in the $J$, $H$ and $K_s$ bands. In PPMXL, data from USNO-B1.0
are used as the first epoch images and those from 2MASS as the second epoch images, deriving the mean positions and proper motions for 910,468,710 objects from the brightest magnitudes down to $V \approx 20$\,mag \citep{Roeser10}. Mean errors of the proper motions vary from $\sim 4$ milli-arcseconds per year (mas/yr) for $J<10$\,mag to more than 10 mas/yr at $J>16$\,mag. 

The field of Praesepe is also covered in the SDSS DR9. A cross-matching between SDSS and PPMXL shows that SDSS is $\sim 95\%$ complete within 5\,deg from from the centre of Praesepe for stars with $J\leq15.5$\,mag but does not contain stars beyond $5.5^\circ$ from the centre of Praesepe. SDSS data covers five optical bands ($ugriz$) to a depth of $g\sim 23$\,mag \citep{York}.

\section{Membership determination} \label{sec:memdet}
\subsection{Astrometric membership}
We restrict our study to a radius of 3.5 degrees from the centre of Praesepe ($\alpha=8^h 40^m 00^s$, $\delta=19^\circ 30' 00''$ \citealt{Lynga}). The reason of this choice is that the tidal radius of Praesepe is $r_t=3.5^\circ\pm0.1^\circ$ \citep{Kraus} suggesting that stars beyond this point are more likely to be background stars. We present our estimation of the background contamination within this radius in section \ref{subsec:SurfaceDensity}.

Within the search area, cluster members are selected based on proper motions followed by two photometric tests, both of which are 
described in more detail further below.
We restrict ourselves to stars with $J < 15.5$\,mag whose proper motion errors are about $\sim 4$\, mas/yr. For fainter stars the proper motion errors become much larger than 10 mas/yr.
A comparison of PPMXL with \citet{Boudreault12} shows that PPMXL is about 93\% complete down to $J=15.5$\,mag, but becomes incomplete 
for fainter magnitudes.

We first select cluster members based on their proper motions.  We use a $\chi^2$ test to separate cluster members from field stars,
 i.e. for each star $i$, we calculate a $\chi^2$ value according to:

\begin{equation} 
 \label{chi2}
 \chi_i^2 = \frac{(\mu_{\alpha i}-\bar{\mu}_{\alpha})^2}{e_{\alpha i}^2+e_{\alpha}^2+\sigma_{\alpha}^2} + 
 \frac{(\mu_{\delta i}-\bar{\mu}_{\delta})^2}{e_{\delta i}^2+e_{\delta}^2+\sigma_{\delta}^2} < 6.17
\end{equation}

\noindent where $\mu_{\alpha i}$ and $\bar{\mu}_\alpha$ are the proper motion of star $i$ and the average cluster proper motion in right ascension,
$\mu_{\delta i}$ and $\bar{\mu}_\delta$ are the proper motion of star $i$ and the average cluster proper motion in declination,
$e_{\alpha i}$, $e_\alpha$,  $e_{\delta i}$, $e_\delta$ are the corresponding errors and $\sigma_\alpha$ and $\sigma_\delta$ are the 
components of the internal velocity dispersion. For the mean cluster motion and the corresponding errors we use $\bar{\mu}_{\alpha}=-35.81$ mas/yr,
 $\bar{\mu}_{\alpha}=-12.85$ mas/yr, $e_{\alpha}=0.29$ mas/yr and $e_{\delta}=0.24$ mas/yr 
as determined by \citet{van Leeuwen}. We also assume a one-dimensional internal velocity dispersion of $\sigma_v=0.67\pm0.23$ km/sec \citep{Madsen}, which corresponds to a proper motion of  $\sigma_v=0.78$  mas/yr at the 
distance of Praesepe. As the threshold separating members from non-members we assume $\chi^2=6.17$, equivalent to a 95.4\% (1.5 sigma) limit for two independent
degrees of freedom. Table \ref{tab:par} summarizes the parameters for Praesepe.

\begin{table}
\caption{Praesepe parameters. ($\alpha_0$,$\delta_0$): cluster centre (see Sec. \ref{sec::center}) ; ($\mu_\alpha$,$\mu_\delta)$: mean proper motion of the cluster; ($e_\alpha$, $e_\delta$): proper motion mean error; $\sigma_v$: one-dimensional internal velocity dispersion}
\begin{tabular}{ccc} 
  \hline
 parameter & value & reference \\ 
 \hline
 \hline
 $\alpha_0$  & \ $08^h 39^m 37^s$ & Our work, 2013 \\
 $\delta_0$ & $+19^\circ 35' 02''$ &  \\ \\
 $(m-M)_0$ & $6.30\pm0.07$\,mag & \citet{van Leeuwen} \\
 $\mu_\alpha$ & $-35.81$\,mas/yr&  \\
 $e_\alpha $ & $0.29$\,mas/yr&  \\
 $\mu_\delta $ & $-12.85$\,mas/yr&  \\
 $e_\delta$ & $0.24$\,mas/yr&  \\ \\
 $\sigma_v$  &  $0.67\pm0.23$ km/sec& \citet{Madsen} \\ \\
 $\log{(\rm age)}$ & $8.77\pm0.1$ dex & \citet{Fossati} \\
 $E(B-V)$ & $0.027\pm0.004$ mag &\citet{Taylor} \\
 $[\rm Fe/\rm H]$ & $0.11\pm0.03$ dex & \citet{An} \\
\hline
 \end{tabular}
\label{tab:par}
\end{table}

\begin{figure}
\centering
\includegraphics[width=84mm]{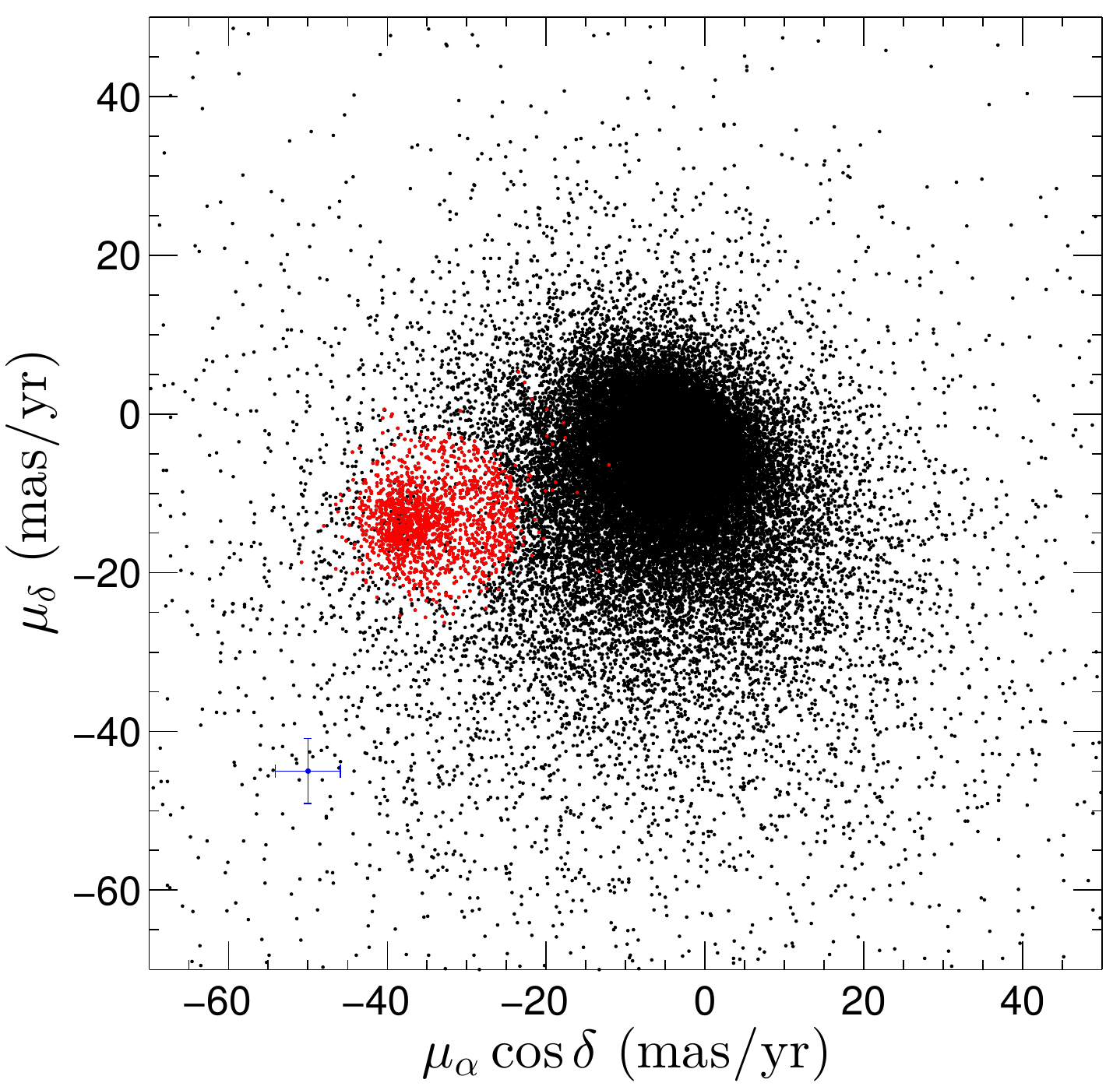}
\caption{Plot of the proper motions of all PPMXL stars in a field of $3.5^\circ$ from the centre of Praesepe (black dots) and the stars that satisfy the proper motion 
test described by Eq. \eqref{chi2} (red dots). The mode of the proper motion errors is shown by the blue cross in the lower left corner.}

\label{fig:PMP}
\end{figure}

We ignore stars with very large proper motion errors ($e_\alpha>15$ mas/yr) in PPMXL since a separation into
field and cluster stars is not possible for them given the absolute value of the proper motion of Praesepe (see also Fig. \ref{fig:PMP}).
We ignore projection effects due to the different location of stars on the sky since they amount to only 0.07 mas/yr difference in proper motion, 
significantly smaller than the error bars in PPMXL.
We also ignore distance effects due to the different radial distances of stars along the light of sight. This distance
effect only adds an uncertainty of about 2 mas/yr which is small compared to the typical proper motion error of PPMXL, although it could be important for bright stars.

Figure \ref{fig:PMP} shows the proper motions of all PPMXL stars in a field of $3.5^\circ$ in radius.
Stars that pass our kinematic test are shown in red. From 45870 stars with $J \leq 15.5$\,mag and $e_\alpha,e_\delta\leq10$ mas/yr that reside within $3.5^\circ$ from the 
cluster centre, 1613 stars meet our $\chi^2$ criterion.

\begin{table*} 
\caption{Celestial coordinates, proper motions, magnitudes and the masses of the possible members of Praesepe sorted by right ascension. Table 2 is available in its entirety in the electronic version of the journal.}\label{tab:members} 
\begin{tabular}{ccccccc} 
\hline 
RA (J2000) & Dec. (J2000) & $\mu_{\alpha}\cos{\delta}$ & $\mu_{\delta}$ & $J$ & $K_{\rm s}$ & Mass \\ 
(deg) & (deg) & (mas/yr) & (mas/yr) & (mag) & (mag) & ($\rm M_\odot$) \\ 
\hline 
126.423399 & +19.511820 & $-33.0\pm4.2$ & $-14.0\pm4.2$ & $15.066\pm0.039$ & $14.168\pm0.048$ & 0.183 \\ 
126.572703 & +19.726598 & $-36.2\pm4.2$ & $-13.7\pm4.2$ & $13.896\pm0.025$ & $12.981\pm0.024$ & 0.327 \\ 
126.597027 & +19.551072 & $-35.6\pm4.2$ & $-22.8\pm4.2$ & $14.643\pm0.032$ & $13.696\pm0.034$ & 0.225 \\ 
... & ... & ... & ... & ... & ... & ... \\
133.536124 & +18.477437 & $-27.1\pm3.7$ & $-08.6\pm3.7$ & $11.910\pm0.023$ & $11.081\pm0.020$ & 0.675 \\ 
133.541163 & +19.946329 & $-26.6\pm3.7$ & $-12.8\pm3.7$ & $12.233\pm0.020$ & $11.385\pm0.021$ & 0.618 \\ 
133.580451 & +19.718509 & $-36.0\pm3.7$ & $-21.4\pm3.7$ & $11.090\pm0.022$ & $10.452\pm0.018$ & 0.825 \\ 
\hline 
\end{tabular} 
\end{table*} 

\subsection{Photometric membership} \label{subsec:phmem}
We use the latest version of the PADOVA stellar evolution models\footnote{\url{http://stev.oapd.inaf.it/cgi-bin/cmd}} from \citet{Marigo} and \citet{Girardi} to estimate the colours and magnitudes of the cluster members based on their metallicity, age and extinction. We use the $J$ and $K_{\rm s}$ bands of the PADOVA models to create an isochrone. 
In addition to PADOVA isochrones, we also use two more isochrones from \citet{Hauschildt} (NextGen models) and \citet{Allard} (BT-Settl models) for comparison. In our study, we adopt an age of 590 Myrs \citep{Fossati}, an extinction of $E(B-V)=0.027$ mag \citep{Taylor} and a distance modulus of $(m-M)_0 = 6.30\pm0.07$\,mag \citep{van Leeuwen} for Praesepe.
 A number of values have been found for the metallicity of Praesepe: $+0.13\pm0.007$\,dex \citep{Boesgaard88}; $+0.125\pm0.032$\,dex \citep{Boesgaard89}; 
[Fe/H] = $+0.038\pm0.039$\,dex \citep{Friel}; $0.11\pm0.03$\,dex from spectroscopy and $+0.20\pm0.04$\,dex from photometry \citep{An} and $+0.27\pm0.10$\,dex \citep{Pace}. Since the location of stars in the CMD is not very sensitive to the adopted metallicity, we only
consider [Fe/H]=$0.11\pm0.03$\,dex \citep{An}.

\begin{figure}
\centering
\includegraphics[width=84mm]{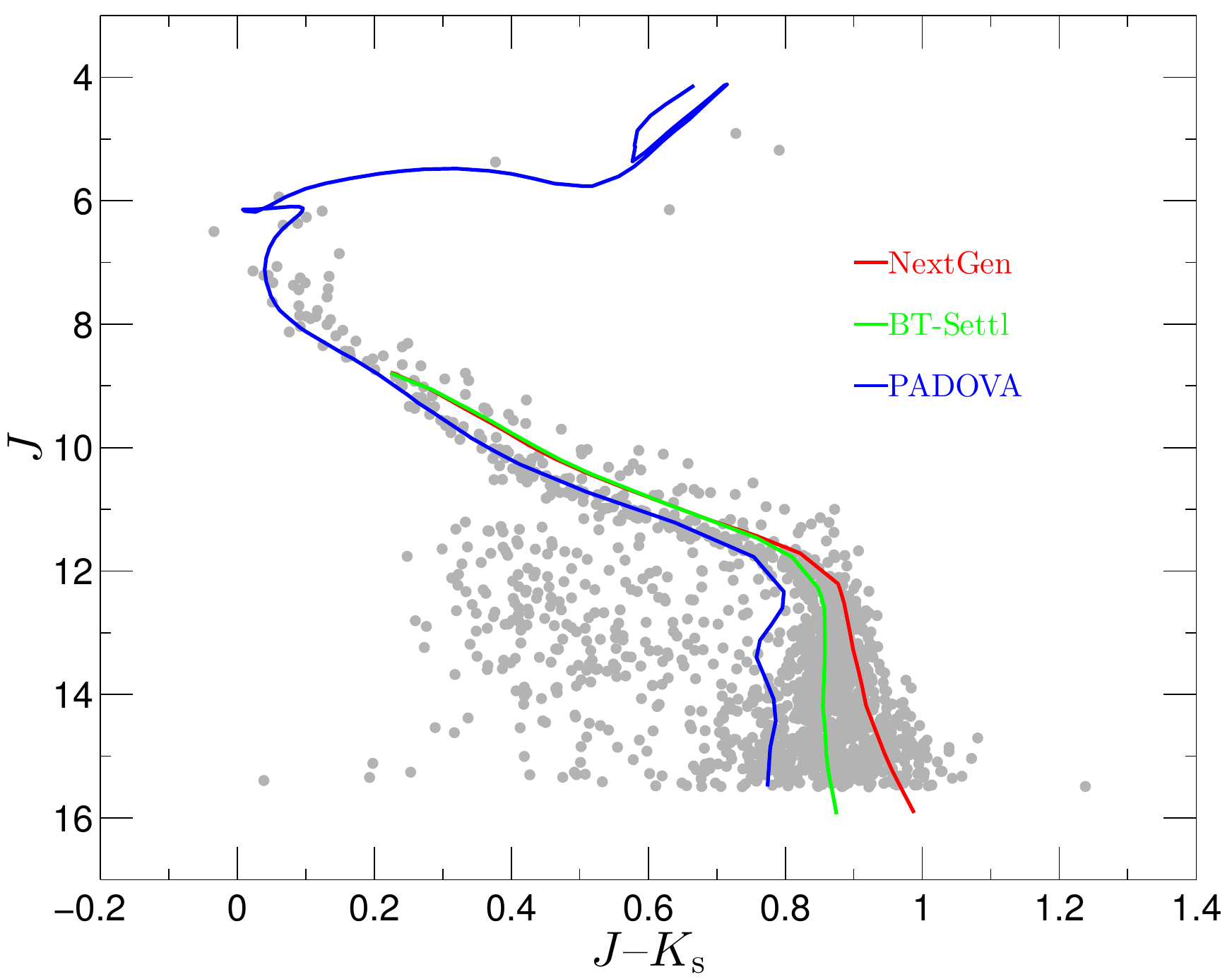}
\caption{Colour-magnitude diagram of the stars selected by proper motions (filled grey circles). 
Solid blue, green and red lines show the PADOVA (left), BT-Settl (middle) and NextGen (right) isochrones respectively.}
\label{fig:CMDPM2}
\end{figure}

Fig. \ref{fig:CMDPM2} shows the distribution of the stars that pass the kinematic test in a CMD as well as the corresponding 
PADOVA (T = 590 Myr, [Fe/H]=0.11\,dex), NextGen and BT-Settl (both with T = 590 Myr, [Fe/H]=0\,dex) isochrones for comparison. We require that photometric members 
lie within $2.5\sigma$ of the isochrones in the $J$ vs $J-K_{\rm s}$ colour magnitude diagram where $\sigma$ refers to the mean error of photometry for each star. It can be seen 
from Fig. \ref{fig:CMDPM2} that the proper motion selected stars agree reasonably well with the PADOVA isochrone down to $J=11.75$\,mag, which corresponds to about 0.7\,M$_\odot$. However, starting
from $J=11.75$\,mag, the PADOVA isochrone disagrees with the actual location of the possible Praesepe
members in the CMD as the isochrone predicts faint (low-mass) stars to be bluer than observed.
Since this disagreement is virtually insensitive to the assumed age, metallicity
and distance modulus, we reason that this is due to the inherent limitations in the PADOVA isochrone
for low-mass stars. Similar results have been obtained by \cite{Roeser11} for the Hyades.

\begin{figure}
\centering
\includegraphics[width=84mm]{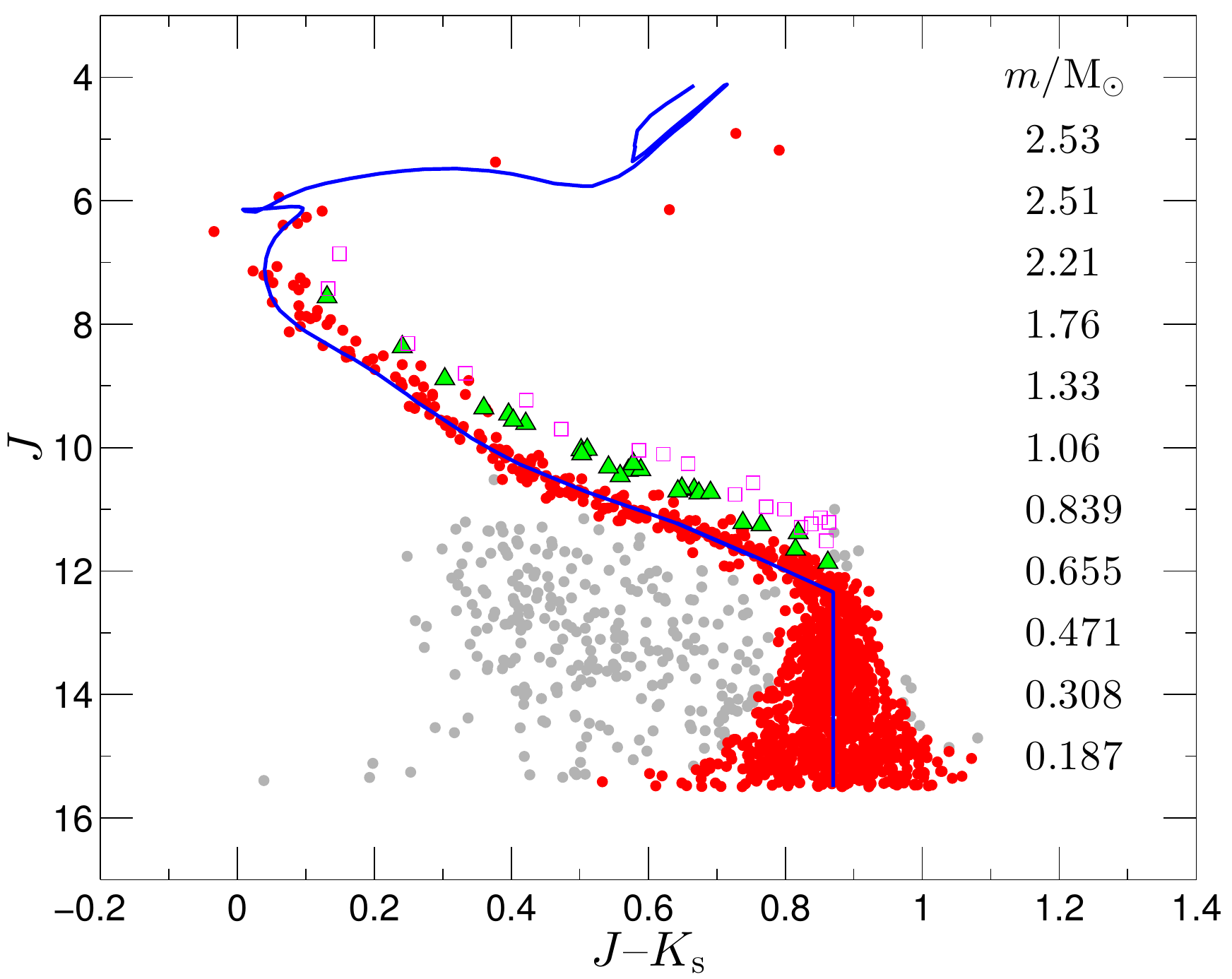}
\caption{Colour-magnitude diagram of the stars selected by proper motion and photometry. Filled red circles show stars that satisfy both tests (1286 stars). 
The filled grey circles show stars that do not pass the photometric test. Inspection of the spatial position of these stars shows that they are evenly spread across the survey region and the distribution of their $\chi^2$ values shows no clustering around zero, hence these stars are most likely field stars. The green-filled triangles show possible binaries (25 stars) and open magenta squares show possible multiples (18 stars). The blue line denotes the modified PADOVA isochrone.}
\label{fig:CMDPMF}
\end{figure}

\begin{figure}
\centering
\includegraphics[width=84mm]{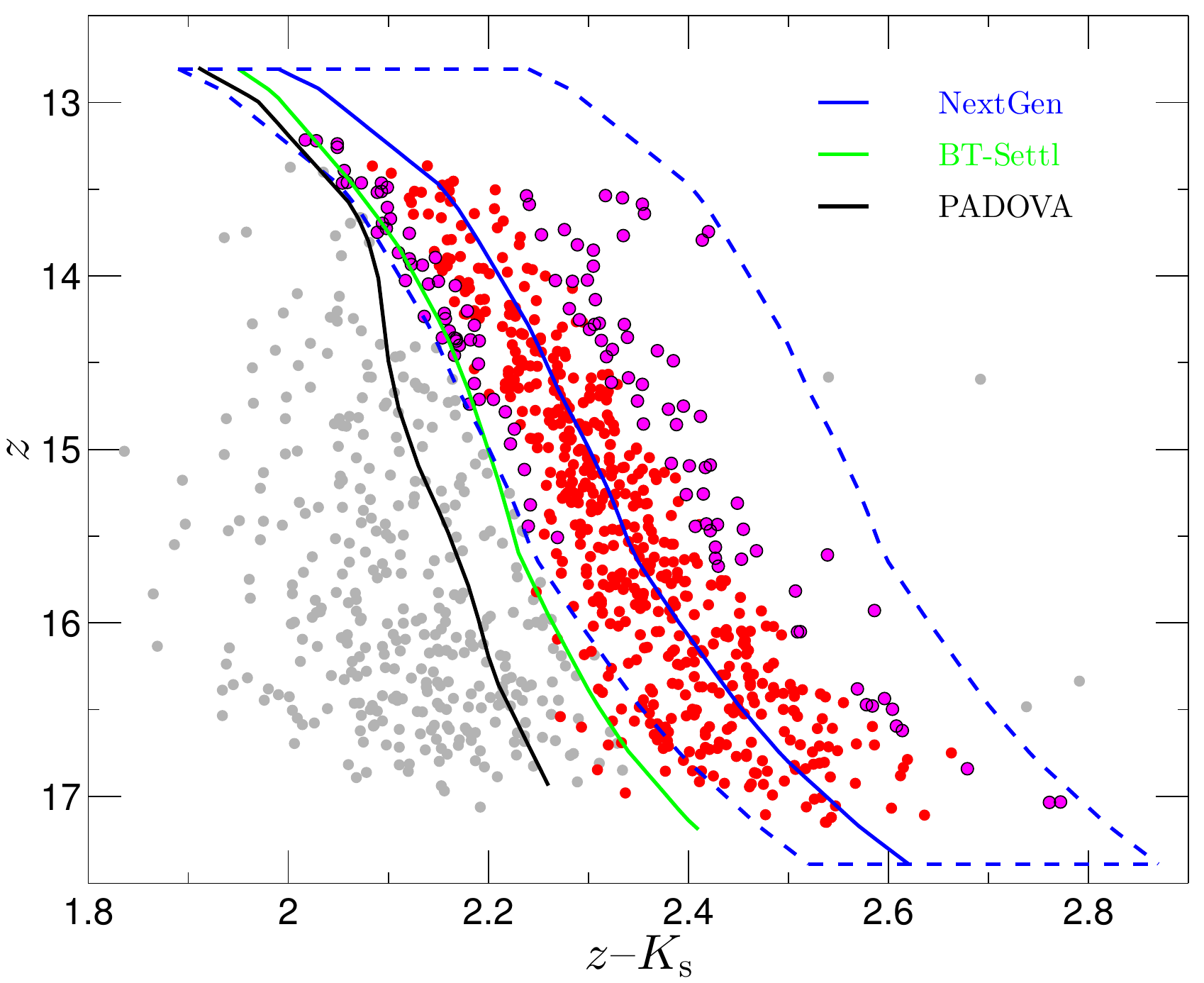}
\caption{Colour-magnitude diagram of all stars with $J>12$ which are selected by the first photometric test (1020 stars). Filled red circles show stars lie within $2.5\sigma$
of the isochrone (528 stars). The filled magenta circles show those stars which are designated as candidate members due to their spatial position and $\chi^2$ distribution of proper motions (124 stars).
The filled grey circles show the stars that fail the second photometric test, hence they are assumed as non-members (368 stars). PADOVA (left), BT-Settl (middle) and NextGen (right) isochrones are represented by black, green and solid blue lines respectively.}
\label{fig:CMDSDSS}
\end{figure}

In order to select faint cluster members photometrically, we shift the PADOVA isochrone by $+0.1$\,mag for magnitudes fainter than $J=11.75$\,mag. This seems better than using the BT-Settl or NextGen isochrones since these isochrones are still a bit off at faint magnitudes and they are off at bright magnitudes as well. 

Fig. \ref{fig:CMDPMF} depicts the stars that satisfy both the proper motion and photometric requirements (filled red circles). 
The blue line is the modified isochrone which determines the photometric membership of the PM-selected members and it matches the average location of the cluster members very well. 
The green-filled triangles and open magenta squares show the possible binary and multiple stars which are discussed in section \ref{subsec:binary}. 

\begin{figure}
\centering
\includegraphics[width=85mm]{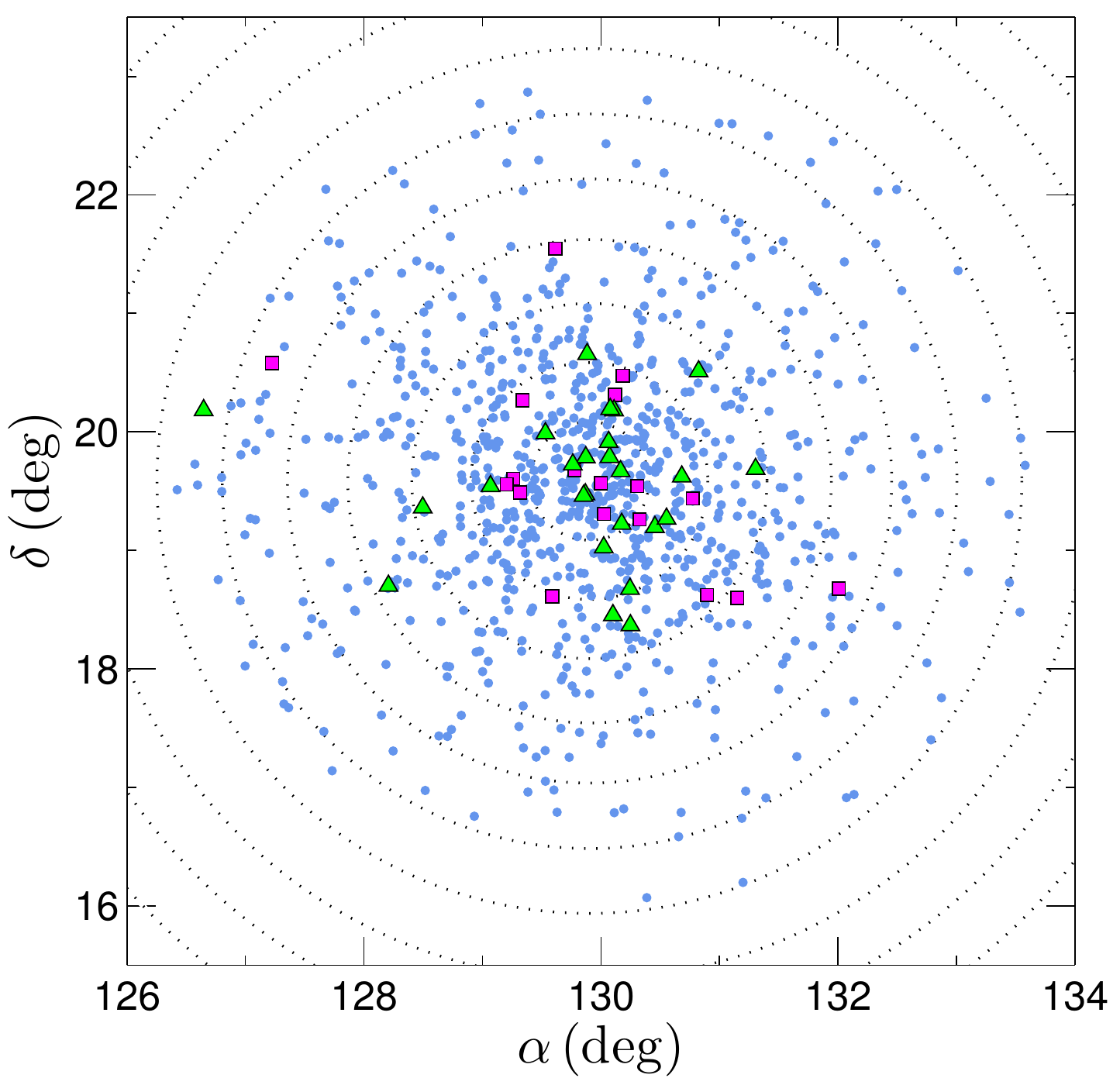}
\caption{Spatial distribution of the possible cluster members of Praesepe (893 stars) with respect to the new centre of Praesepe. The filled green triangles and the filled magenta squares show possible binaries and multiples. Dotted grey circles are spaced by $0.5^\circ$.}
\label{fig:SpaceDist}
\end{figure}

There are 1286 stars that satisfy both tests, however, due to the relatively large errors of proper motion and photometry of faint stars we expect a significant contamination of field stars for magnitudes $J>12$\,mag.
In order to remove possible contaminants with $J>12$\,mag we apply a second photometric test using $z$ magnitudes from SDSS and $K_{\rm s}$ magnitudes from 2MASS.
Fig. \ref{fig:CMDSDSS} shows the position of all stars with $J>12$\,mag in a $z$ versus $z-K_{\rm s}$ diagram. One can see that in this CMD, the NextGen isochrone matches the actual position of possible members. As a result we use this isochrone to remove background stars photometrically. We require that each photometric member lies within $2.5\sigma$
of the isochrone (filled red circles) or its color index differs by no more than $-0.1/+0.25$\,mag from the corresponding color index of the isochrone at the same magnitude (filled magenta circles). This additional condition is added since we observe that there are a number of stars (enclosed by the dashed blue lines in Fig. \ref{fig:CMDSDSS}), which are not identified as photometric members due to their small photometric errors but are concentrated towards the cluster centre and have a $\chi^2$ distribution clustered around zero. As a result we keep these stars as photometric members. In contrast, the stars shown as filled grey circles in Fig. \ref{fig:CMDSDSS} have a flat $\chi^2$ distribution and are uniformly spread in our 3.5-degree survey field. We end up with 893 stars which satisfy the astrometric and all photometric tests. Figure \ref{fig:SpaceDist} shows the spatial distribution of these stars. We also obtain the mass of the possible members by interpolating the theoretical masses from the PADOVA isochrone using the $J$ magnitudes. These masses are listed in Table \ref{tab:members}. 

\subsection{Cluster centre}\label{sec::center}
Before obtaining surface density plots, we first determine a new coordinate for the cluster centre (density centre). 
The density centre as defined by \citet{von Hoerner6063} is the density weighted average of the positions of all stars:

\begin{equation}\label{Eq:rden}
\vec{r}_{d,j}\equiv\frac{\sum_{i}{\vec{r}_i\rho^{(i)}_j}}{\sum_{i} \rho^{(i)}_j}
\end{equation}

\noindent where $\rho^{(i)}_j$ is the local density estimator of order j around the $i$th particle with position vector $\vec{r}_i$. We replace the 3D density estimator $\rho^{(i)}_j$ by the surface density $\Sigma^{(i)}_j$. To estimate $\vec{r}_{d,j}$ we adopt the unbiased form of the density estimator introduced by \citet{Casertano} and consider the 10 nearest neighbours of each star to obtain $\Sigma^{(i)}_j$.  We then obtain the new coordinate of the cluster centre as follows:
$$\alpha_{\rm centre}=8^h 39^m 37^s \quad \delta_{\rm centre} = 19^\circ 35' 02''$$

This new centre differs by about $5''$ from the coordinate of the cluster center given by \cite{Lynga} ($\alpha_{\rm centre}=8^h 40^m 00^s, \delta_{\rm centre} = 19^\circ 30' 00''$).

\subsection{Surface density profile} \label{subsec:SurfaceDensity}
To determine the distribution of member stars and the possible background stars, we first consider a field
of $5^\circ$ and divide this field into 18 radial bins from $0^\circ$ to $5^\circ$ where SDSS is $\sim95\%$ complete.

\begin{figure}
\centering
\includegraphics[width=85mm]{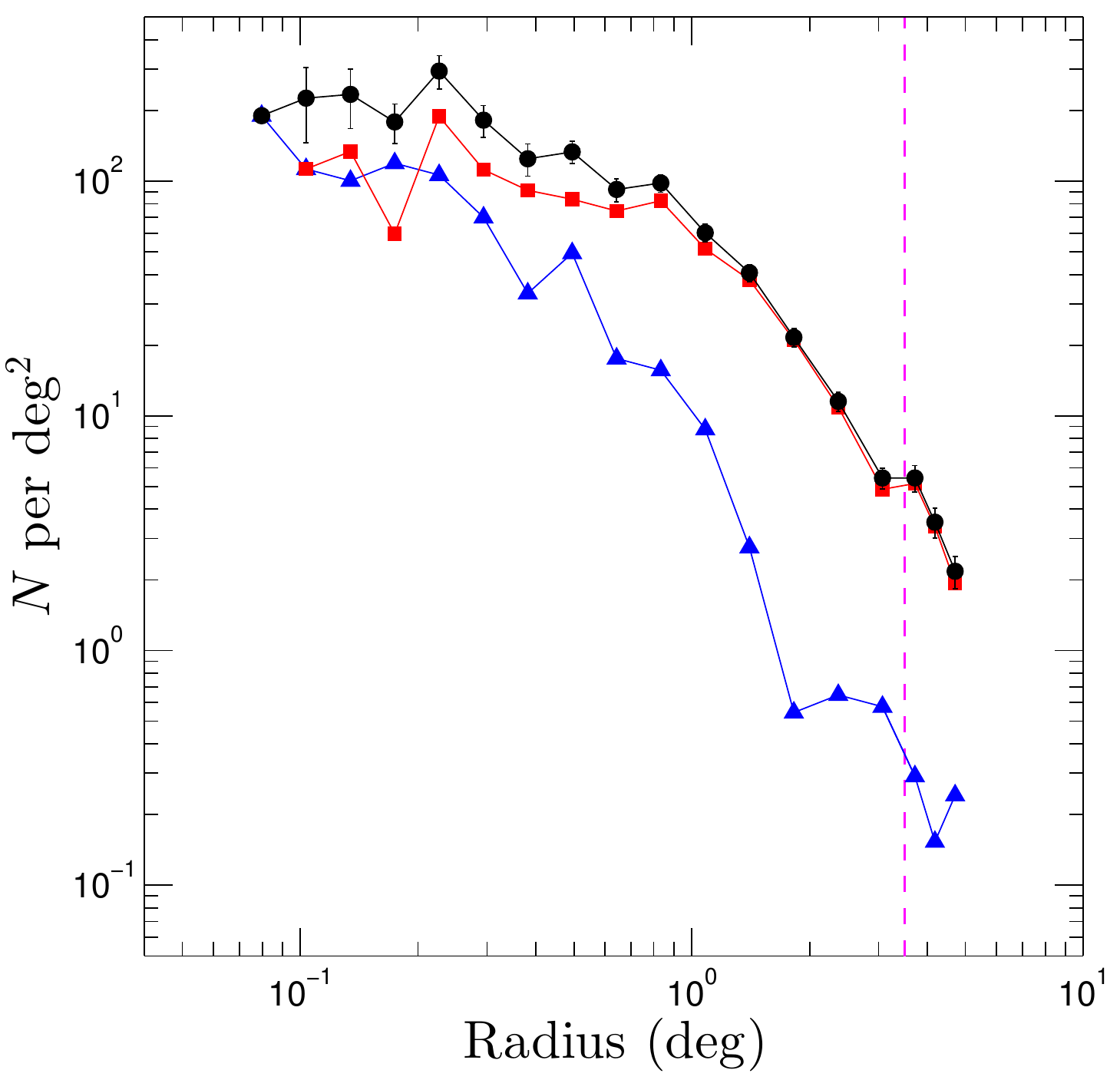}
\caption{Surface density of all ($J\leq15.5$, upper line), faint ($11<J\leq15.5$, middle line) and bright ($J\leq11$, lower line) possible members in a 5-degree field around the centre of Praesepe. The dashed magenta line marks the tidal radius of Praesepe. To avoid confusion only error bars of the density of all stars are shown. The error bars are Poissonian. }
\label{fig:density1}
\end{figure}

 We then consider all stars that satisfy the astrometric and photometric selection criteria in this field and plot the surface density of faint ($11< J\leq15.5$), bright ($J\leq11$) and all stars ($J\leq15.5$) in each bin as a function of radius from the new centre of Praesepe in Fig. \ref{fig:density1}. At the distance of Praesepe $J=11$\,mag corresponds to $m=0.84$\,M$_\odot$. Error bars only consider the statistical uncertainties in the number of stars. One can see that the density profiles do not level off beyond the tidal radius of Praesepe. This is in agreement with \cite{Kuepper} who showed that the surface density profiles of star clusters extend beyond the tidal radius due to stars escaping the cluster. To estimate an upper limit for the density of background stars, we therefore consider the density of the last radial bin and find that the background density of stars is $2.17\pm0.34$ stars/deg$^2$, implying that there are still about $84$ background stars within a field of 3.5 degrees from the centre of Praesepe. Since we found 893 candidate stars in total within the same area, the background contamination is $\sim 10$\% and needs to be considered when determining the mass function of cluster members and the total cluster mass. We will statistically subtract these background stars when deriving the mass function. The details of this subtraction are explained in Sec. \ref{sec:MF}. 

\begin{figure}
\centering
\includegraphics[width=84mm]{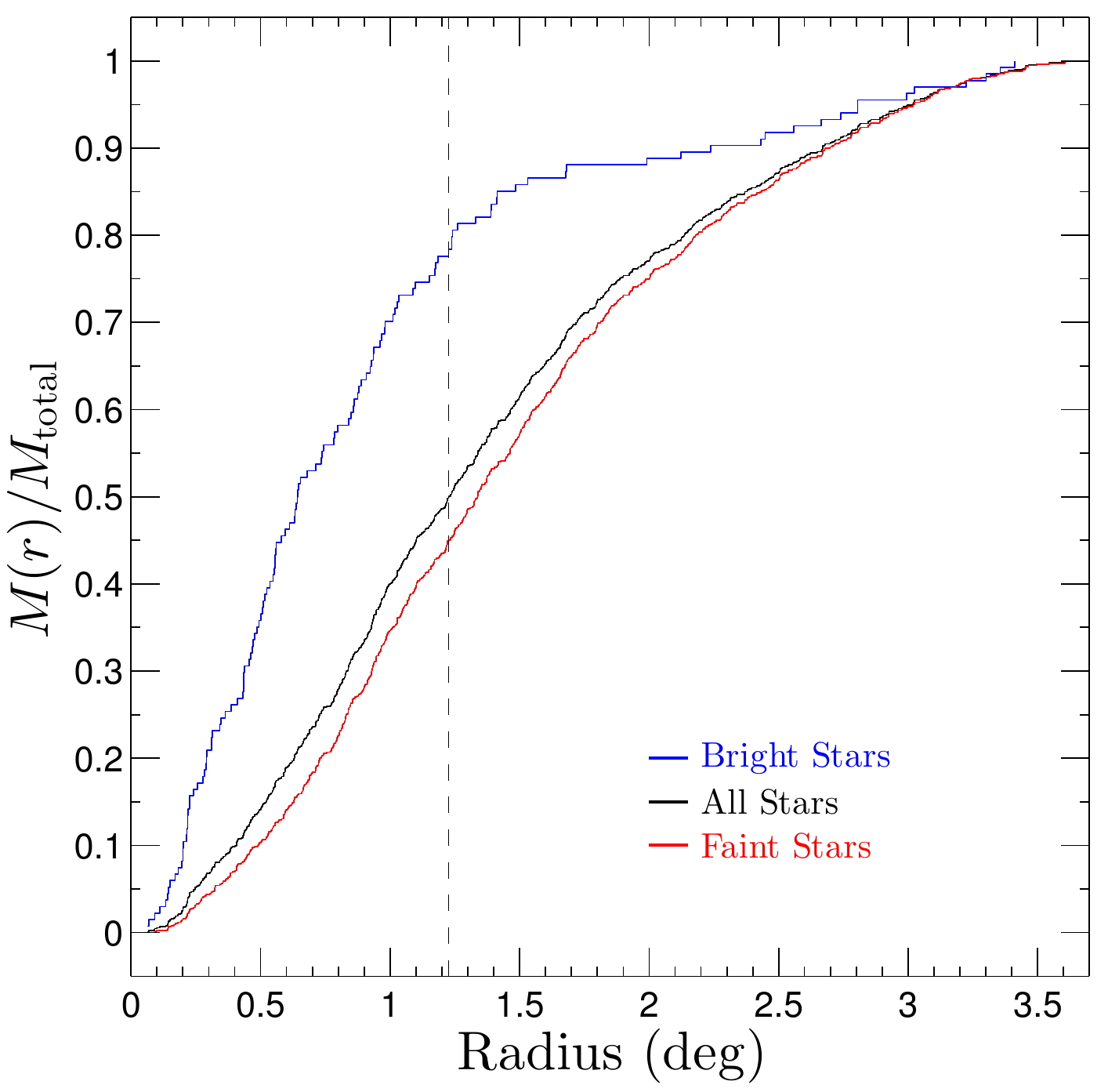}
\caption{Normalized cumulative sums of stellar masses for bright (upper line), faint (lower line) and all (middle line) cluster members as functions of radius in a field of $3.5^\circ$ from the new cluster centre. The dashed line intersects the total cumulative curve at half the maximum value corresponding to a half mass radius of $1.23^\circ$ which is equivalent to 3.90 pc.}
\label{fig:density2}
\end{figure}

\begin{figure}
\centering
\includegraphics[width=84mm]{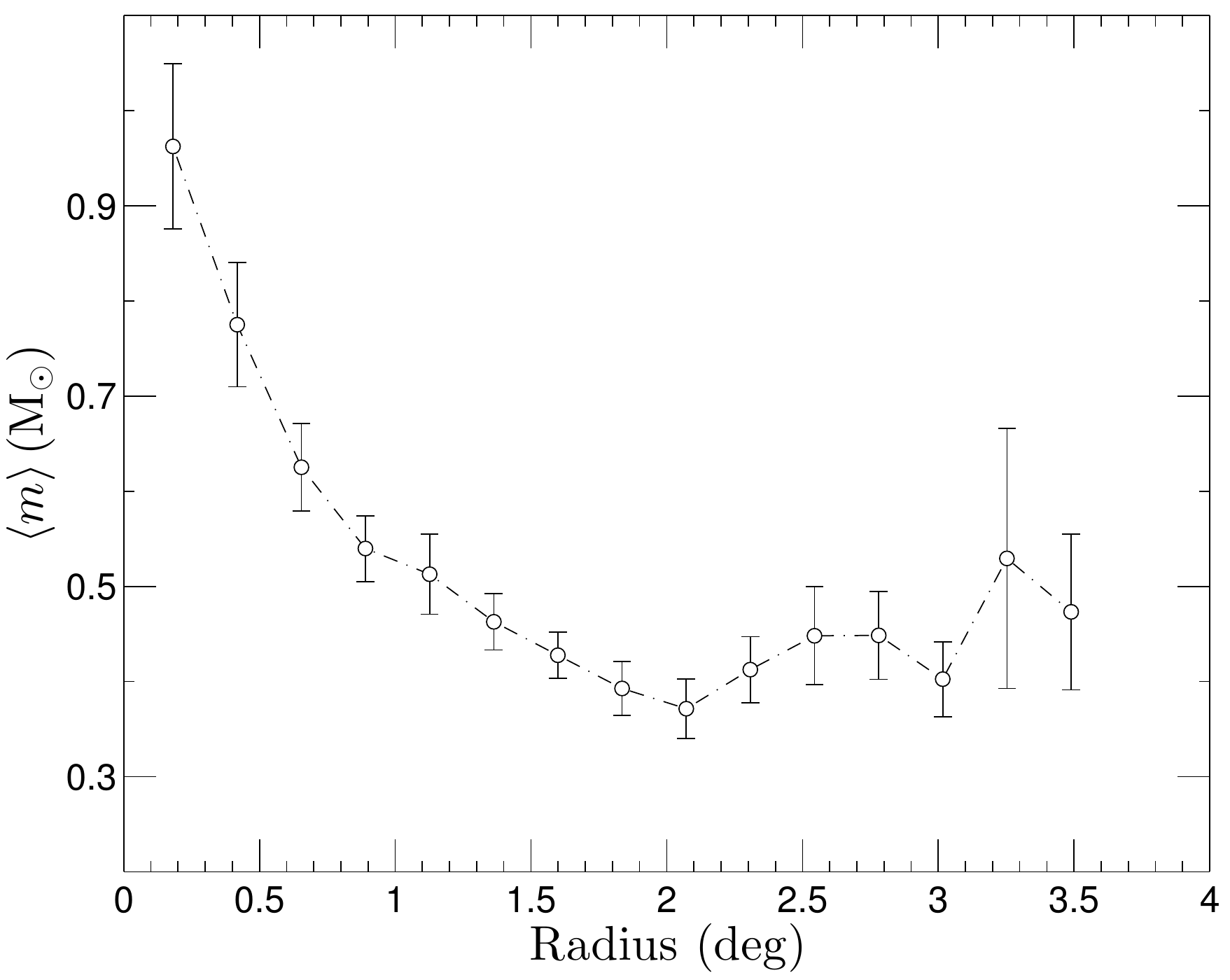}
\caption{The average mass of the possible cluster members as a function of radius from the centre of Praesepe. The average mass is roughly constant beyond $1.5^\circ$.}
\label{fig:avgmass}
\end{figure}

Fig. \ref{fig:density2} shows the cumulative sums of stellar masses for faint, bright and all stars as a function of radius in the search field of $3.5^\circ$. 
The two-dimensional half-number radius and half-mass radius are $1.34^\circ$ and $1.23^\circ$ which correspond to $4.25$\,pc and $3.90$\,pc respectively. 
Assuming that the corresponding 3D radii are 1/3 larger than the projected radii, we find that the 3D half-number and half-mass radii 
are $5.67$ and $5.20$\,pc respectively. Table \ref{tab:par2} summarizes half-mass and half-number
radii corresponding to faint, bright and all stars. The fact that the 2D half-mass and half-number radii of all cluster stars are about 2 times larger than those of bright stars evidently shows that Praesepe is strongly mass segregated. Mass segregation of Praesepe can also be inferred from Fig. \ref{fig:avgmass} which shows the average mass of the possible cluster members as a function of radius from the new cluster centre. One can see that the average mass of stars is $0.96\pm0.08$\,M$_\odot$ inside 0.3\,pc and drops down to $\sim0.45$\,M$_\odot$ outside the half-mass radius.

\begin{table}
\caption{Table of projected half-number and half-mass radii for cluster members}
\begin{tabular}{cccc} 
  \hline
type of stars&number&half-number&half-mass\\
  & of stars & radius (2D) & radius (2D) \\  
 \hline
 \hline
 Faint Stars &759 stars&$1.23^\circ$ & $1.34^\circ$ \\
 $11< J\leq15.5$ & & $3.90$\,pc   &  $4.25$ pc \\
\hline
 Bright Stars &134 stars& $0.65^\circ$ &   $0.64^\circ$ \\
  $J\leq11$ & & $2.07$\,pc &  $2.04$ pc \\
\hline
 All Stars & 893 stars &$1.34^\circ$ & $1.23^\circ$  \\
 $J\leq15.5$ & & $4.25$\,pc & $3.90$\,pc \\
\hline
 \end{tabular}
\label{tab:par2}
\end{table}

\section{Results} \label{sec:results}

\subsection{Stellar mass function}\label{sec:MF}
As shown before, possible cluster members are mainly located within 3.5 degrees of the cluster centre so we restrict our study to stars within this radius when
deriving the mass function. To obtain the mass function of the cluster members, we must statistically subtract the background level. The is done as follows:
\begin{enumerate}
\item In Fig. \ref{fig:density1}, we assume that all stars which reside between $4^\circ$ and $5^\circ$ are backgrounds stars.
\item We derive the mass function of these background stars.
\item We generate 84 random stars distributed according to the mass function derived in the previous step.
\item Using these stars, we select possible members with similar mass and remove them from the list of possible members.
\end{enumerate}

\begin{figure}
\centering
\includegraphics[width=84mm]{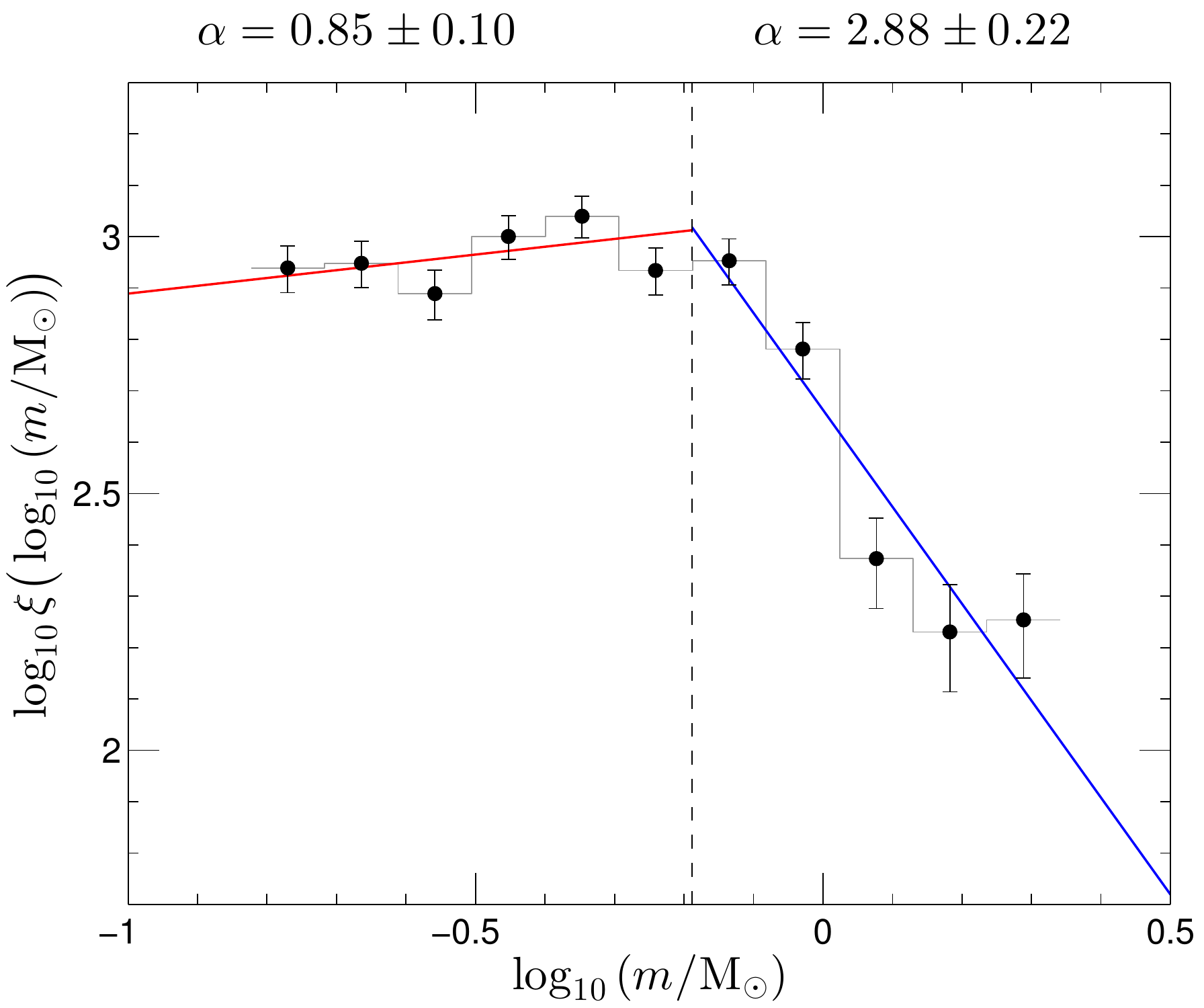}
\caption{Mass function of stars in Praesepe corrected for background stars. 
Error bars of data points are Poissonian. Dashed line which corresponds to $m=0.65$\,M$_\odot$ shows the point at which the mass functions has a turnover. Red ($\alpha=0.85\pm0.10$) 
and blue ($\alpha=2.88\pm0.22$) lines show the best fit to the mass function of low-mass and high-mass stars respectively. While in this figure data is binned, the procedure we use to determine the $\alpha$ values does not work with binned data.}
\label{fig:OverallMF}
\end{figure}

The mass function for the possible members after background subtraction is shown in Fig. \ref{fig:OverallMF}. This figure shows that the mass function of Praesepe
cannot be fitted by a single power-law distribution $dN/dm=\xi(m) \sim m^{-\alpha}$. As a result we assume that the mass function is described by a two-stage power-law distribution with a turnover at $m_{\rm t}$, so that the mass function of stars with masses higher than $m_{\rm t}$, hereafter high mass stars, is fitted by $\xi(m) \sim m^{-\alpha_{\rm high}}$, whereas the mass function of those lower than $m_{\rm t}$, hereafter low mass stars, is fitted by $\xi(m) \sim m^{-\alpha_{\rm low}}$. The method we use to derive $m_{\rm t}$ and the value of each $\alpha$ is based on a maximum-likelihood fitting combined with a Kolmogorov-Smirnov test to evaluate the goodness of the fit. This method was introduced by \cite{Clauset} for the case that the exponent of the power-law $\alpha$ is greater than unity and there exists a lower bound to the data. We re-derive the relations for both $\alpha>1$ and $\alpha<1$ and when the data is bounded on two sides. The case of two-sided bounds is important in our case, since stars more massive than 2.20\,M$_\odot$ have evolved off the main sequence while stars less massive than 0.15\,M$_\odot$ are too faint to be detected. The details of this method and the derivation of $\alpha$ values and their corresponding errors are presented in appendix \ref{sec:powerlaw} . This method provides much more accurate results compared to using a least-squares method on binned data \citep{Clauset}. 

Using the method mentioned above, we find that a K-S test indicates with high confidence (5\% significance level) that the turnover in the mass function is at $m_{\rm t}=0.65$\,M$_\odot$ which is also visible in the figure. We also obtain $\alpha_{\rm low}=0.85\pm0.10$ and $\alpha_{\rm high}=2.88\pm0.22$ for low and high mass stars respectively. According to the K-S test our result for the overall mass function would also be consistent with a turnover at $0.5$\,M$_\odot$ and in such a case the mass functions slopes would be $\alpha_{\rm low}=0.72\pm0.13$ and $\alpha_{\rm high}=2.51\pm0.15$ (close to $\alpha=2.35$ found by \citealt{Salpeter}).

Since the total mass of Praesepe is about $\sim500$\,M$_\odot$ and the 3D half-mass radius is $\sim5.20$\,pc , the age of Praesepe is approximately $\sim 5$ times its relaxation time and one can assume that the stellar content of the cluster has thoroughly mixed in the entire cluster, meaning that the location of the turnover in the mass function should not change from one point to another. We therefore fix the location of the turnover and derive the mass function slopes inside and outside the half-mass radius. We find that the mass function slope of massive stars inside the half-mass radius ($\alpha=2.32\pm0.24$) is less steep than the overall slope ($\alpha_{\rm high}=2.88\pm0.22$). However in the outer radial bin, the mass function of massive stars becomes steeper ($\alpha=4.90\pm0.51$), and this is again indicating mass segregation in Praesepe.

To correct for the effect of unresolved binaries on the observed mass function, we did a series of a simulations which are discussed in Section \ref{subsec:binary}.

\subsection{Comparison to other works}

Table \ref{tab:par3} compares mass function slopes for Praesepe as obtained by various studies within the last 20 years. In this table only those studies whose mass range overlaps our studied mass range are listed. 

\par We obtain $\alpha=0.85\pm0.10$ for stars in the mass range $0.15\leq m/{\rm M}_\odot\leq0.65$. Within the error bars, this value is in good agreement with \citet{Baker} and \citet{Boudreault12}, however it is steeper than the
value found by \citet{Adams}. This disagreement, as discussed by \citet{Boudreault10}, is most likely due to the fact that the membership criterion of \citet{Adams} 
is based on a threshold for the membership probability of only $p=0.01$, which increases the likelihood of contamination by background stars. 
The value that we have found for the low-mass slope of the mass function is less steep than the values found by \citet{Hambly}($\alpha=1.5$), \citet{Kraus}($\alpha=1.4\pm0.2$) and \citet{Boudreault10}($\alpha=1.8\pm0.1$). The discrepancy with \citet{Boudreault10} is due to the fact that they identified the members of Praesepe only through photometry, while \citet{Boudreault12} used proper motions combined with photometry, which led to the rejection of more background stars. \cite{Boudreault12} found 1116 members for Praesepe among which 855 stars have $J\leq15.5$\,mag (our survey limit) and 552 of these stars ($\sim65\%$) are recovered by our analysis.

Hence, three recent independent studies (our work, \citealt{Boudreault12} and \citealt{Baker}) show a turnover at $\sim 0.5\pm0.1$\,M$_\odot$in the mass function and a slope of $\alpha \approx 0.8$ for the mass function before the turnover.  

\begin{table*}
\caption{A comparison of different profiles obtained for global mass function of Praesepe.}\label{tab:par3}
\begin{tabular}{|ccc|}
  \hline
Reference & low-mass MF slope & high-mass MF slope \\ 
 \hline
 \hline
Our work (2013) & $\alpha=0.85\pm0.10$ & $\alpha=2.88\pm0.22$  \\
  & $0.15\leq M/M_\odot \leq0.65$   &  $0.65\leq M/M_\odot \leq2.20$ \\
  corrected for binaries & $\alpha=1.05\pm0.05$ & $\alpha=2.80\pm0.05$ \\
\hline
 \citet{Boudreault12}  & $\alpha=0.63\pm0.11^{a}$& --  \\
  &   $0.062\leq M/M_\odot \leq0.695$ & --  \\
\hline
 \citet{Boudreault10}  & $\alpha=1.8\pm0.1$ & --  \\
  &   $0.1\leq M/M_\odot \leq0.6$ & --  \\
\hline
  & $\alpha=1.10\pm0.37$ & --  \\
&   $(Z$ band$) \ \ 0.125\leq M/M_\odot \leq0.6$ & --  \\
\citet{Baker}  & $\alpha=1.07$ & --  \\
&   $(J$ band$) \ \ 0.20\leq M/M_\odot \leq0.5$ & --  \\
  & $\alpha=1.09$ & --  \\
&   $(K$ band$) \ \ 0.20\leq M/M_\odot \leq0.5$ & --  \\
\hline
 \citet{Kraus}  & $\alpha=1.4\pm0.2$ & --  \\
  &   $0.12\leq M/M_\odot \leq1$ & -- \\
\hline
 \citet{Adams}  & $\alpha=0$ & $\alpha=1.6$  \\
  & $0.1\leq M/M_\odot \leq0.4$   &  $0.4\leq M/M_\odot \leq1$ \\
\hline
 \citet{Hambly}  & $\alpha=1.5$ &   \\
  & $0.1\leq M/M_\odot \leq0.5$   &   \\
\hline
 \end{tabular}
\caption*{a) The value of $\alpha=0.63\pm0.11$ is not reported in \citet{Boudreault12}. We derive this value using the original data given in Table 3 of the corresponding paper.}
\end{table*}

\subsection{Binary fraction}
\label{subsec:binary}

\begin{figure}
\centering
\includegraphics[width=84mm]{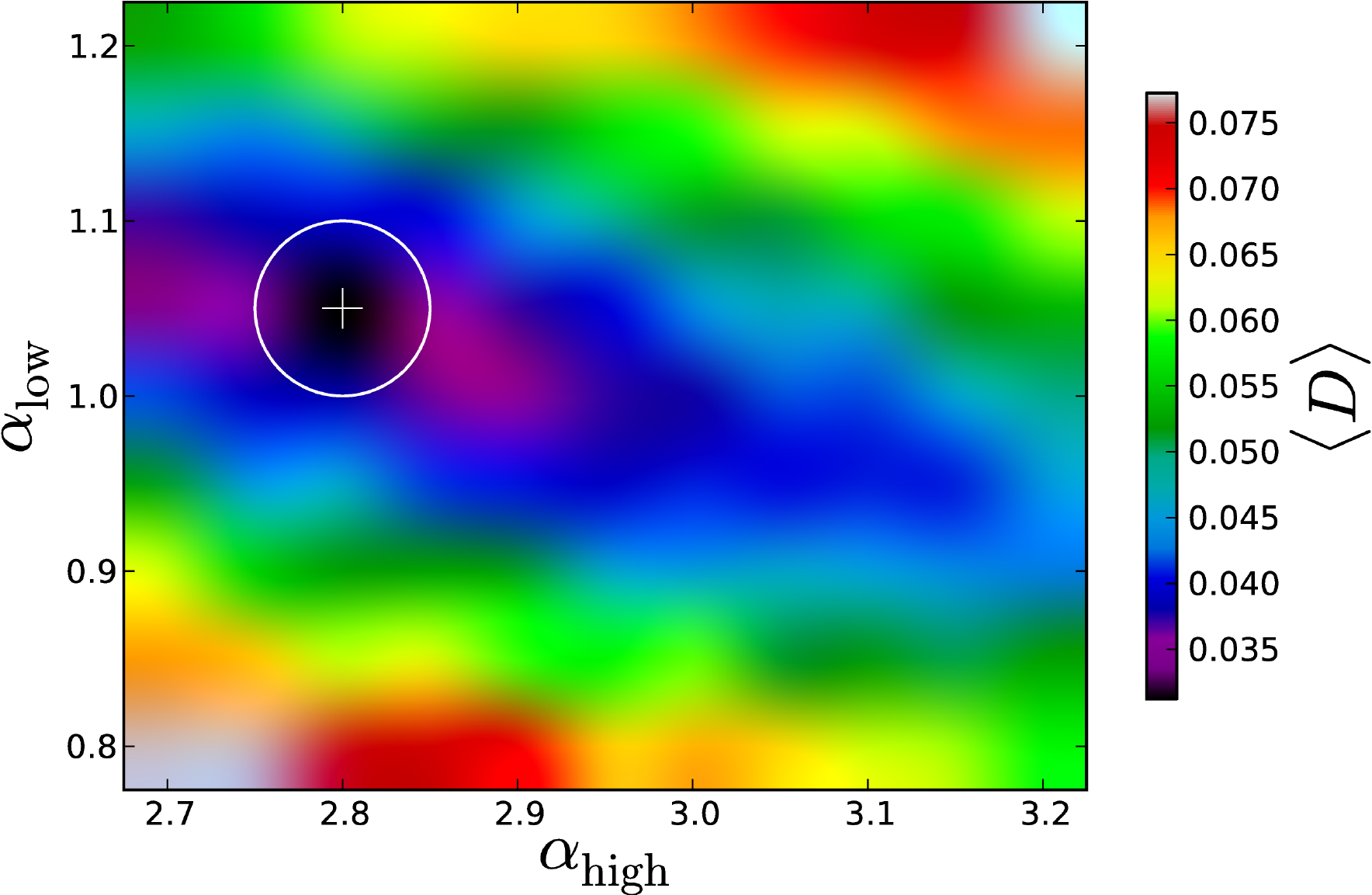}
\caption{The value of the KS statistic $D$ between the outcome of the simulations (averaged over 50 different random seed numbers) and the observed mass function as a function of $\alpha_{\rm low}$ and $\alpha_{\rm high}$ for an assumed binary fraction of $f_{\rm bin}=35\%$. The white cross shows the location of the best match with observation. The white circle represents the lower limit of the errors in our simulation}
\label{fig:simulations}
\end{figure}

Figure \ref{fig:CMDPMF} shows that there are a number of stars which reside above the main sequence of the isochrone (filled green triangles). 
This deviation is significantly outside the error bars for the photometry, so it suggests the presence of binary systems in the cluster. 
The CMD shows that there are also stars (depicted by open magenta squares in Fig. \ref{fig:CMDPMF}) that are more than 0.75 mag above the isochrone.
Since these stars show a strong concentration towards the cluster centre as shown in Fig \ref{fig:SpaceDist}, they are most likely cluster members and are therefore either 
binaries or higher order multiple systems.\footnote{By higher order multiples we mean triple, quadruple and more complex systems.}

There are $25$ stars that deviate from the isochrone by more than $2.5\sigma$, where $\sigma$ refers to the mean error of the photometry.
The corresponding binary fraction, which we define as the fraction of binary systems $f_{\rm bin}= N_{\rm bin}/(N_{\rm sing}+N_{\rm bin})$ is therefore $8.47\%\pm1.55\%$ for $7.05\leq J\leq12.35$. 

In order to recover the true binary fraction and the true mass function, we simulate the effect of binaries through a number of 
Monte Carlo simulations by constructing non-random pairs out of the single stars, assuming that the likelihood of each star to be chosen as a
primary increases linearly with the logarithm of its mass, similar to what is seen in recent simulations of star formation (e.g. \citealt{Bate}).

The procedure that we have adopted to derive the true binary fraction and mass function in our simulations is as follows:

\begin{enumerate}
\item Create a model cluster using a two-stage mass function with a mass function slope for low-mass stars $\alpha_{\rm low}$ and a slope for high-mass stars $\alpha_{\rm high}$.
\item Derive $J$ and $K_{\rm s}$ magnitudes of the stars using the modified PADOVA isochrone.
\item Select each primary component of a binary with a probability which is proportional to $\log_{10}(m_{\rm prim}/m_{\rm min})$.
Here $m_{\rm prim}$ is the mass of each primary component and $m_{\rm min}$ is the minimum mass of all stars in the mass function.
\item Select each secondary component ($J_2$, $K_{s2}$) based on a uniform (flat) distribution for mass ratios ($q=m_s/m_p$) between $0<q<1$.
\item Compute the total magnitude of the binary system: 
\begin{align}
J_{\rm (binary)}&=-2.5\log(10^{-0.4J_1}+10^{-0.4J_2}) \\
K_{\rm s(binary)}&=-2.5\log(10^{-0.4K_{s1}}+10^{-0.4K_{s2}})
\end{align}
\item Add synthetic photometric errors to the magnitudes derived in the previous section. The synthetic photometric errors are generated in such way that they replicate the photometric errors of stars in PPMXL.
\item Determine the binary systems that deviate significantly from the isochrone and count them as observed binaries.
\item Compare the observed binary fraction in the simulation with the binary fraction from the simulation and the mass function of the single stars and non-detected binaries 
   with the observed mass function.
\item Change the parameters of the initial mass function and the assumed binary fraction until the best match in terms of the KS statistic $D$ with the observed mass function and the observed binary fraction is found.
\end{enumerate}

The range of parameters that we use in our simulations to generate the model clusters is:
\begin{itemize}
\item $\alpha_{\rm low}\in [0.8:0.05:1.2]$
\item $m_t=0.65$\,M$_\odot$
\item $\alpha_{\rm high}\in[2.7:0.05:3.2]$
\end{itemize}

In order to reduce statistical random errors we run our simulation for 50 different random seed numbers and then take an average over the KS statistic $D$ and $f_{\rm bin}({\rm observed})$ for each set of given parameters.

\par According to our simulations, an assumed binary fraction of $f_{\rm bin}=35\%\pm5\%$ in the mass range $0.6 \leq m/\rm M_\odot \leq 2.20$ ($7.05\leq J\leq12.35$) and an initial mass function with $\alpha_{\rm low}=1.05\pm0.05$ and $\alpha_{\rm high}=2.80\pm0.05$ yields the best agreement with the observed mass function and observed binary fraction. The fact that $\alpha_{\rm low}$ which is obtained from simulations is steeper than than the observed (uncorrected) value is to be expected since many low-mass stars will be hidden in binaries with more massive companions.

\par Fig. \ref{fig:simulations} shows the value of KS statistic $D$ as a function of $\alpha_{\rm low}$ and $\alpha_{\rm high}$ for $f_{\rm bin}=35\%$. The best match with the observation is obtained by minimizing the value of $D$ and its location in the parameter space is shown by the white cross. As shown in the figure, one can see that the value of $D$ strongly depends on $\alpha_{\rm low}$ in contrast to $\alpha_{\rm high}$ which does not change the value of $D$ significantly. 

Our simulations also reveal that due to photometric errors, a fraction of binaries scatter above the isochrone by more than $\Delta J=0.75$\,mag and 
are identified as fake multiples. According to our simulations, from 18 multiples detected in the CMD, $7\pm2$ can be explained by binaries this way. 
The rest are likely to be genuine multiples.

For comparison, \citet{Bouvier} found a binary fraction of $25.3\pm5.4\%$ for 149 G and K dwarfs observed in Praesepe using adaptive optics. According to our best fitting model, we find a binary fraction of $30\pm5\%$ for the same mass range which is roughly consistent with the results of \citet{Bouvier}.

Finally, we estimate the total cluster mass. By summing up the masses of members obtained from our modified 
PADOVA isochrone and subtracting the contribution of contaminants we derive a total mass of 424\,M$_\odot$. By extending the global mass function 
profile to 0.08\,M$_\odot$ to compensate for missing low mass stars, 
this value increases to 448\,M$_\odot$. Assuming that there are no black holes or neutron stars left in the cluster, we also extend the 
global mass function profile to an initial mass of 8\,M$_\odot$ to calculate the contribution of white dwarfs to the total mass. We adopt the semi-empirical 
relation from \citet{Kalirai} which gives the final mass of white dwarfs as a function of the initial masses of the main-sequence progenitors. After this 
correction, the total mass is 465\,M$_\odot$. Due to binaries, which increase the mass by approximately 
a factor of 0.35, we derive a total mass of $\approx 630$\,M$_\odot$ for Praesepe.

\section{Conclusion}
\label{sec:conclusion}
We have identified 893 possible members of the open cluster Praesepe using proper motions from the PPMXL catalogue 
and $J$ and $K$ photometry from the 2MASS and $z$ photometry from SDSS in a field of $3.5^\circ$ from the cluster centre. 

\par We then calculated a new density centre for the cluster as defined by \citet{von Hoerner6063} using the unbiased form of the local density
estimator from \citet{Casertano}. Using this new cluster centre ($\alpha_{\rm centre}=8^h 39^m 37^s , \delta_{\rm centre} = 19^\circ 35' 02''$) we derived the surface density profile, 2D half-number ($4.25$\,pc) and half-mass radii ($3.90$\,pc) for Praesepe. We found that Praesepe is strongly mass segregated.

\par We derived the global and radial mass functions of Praesepe. The global MF of Praesepe is a two-stage power law with $\alpha_{\rm low}=0.85\pm0.10$ and $\alpha_{\rm high}=2.88\pm0.22$ for low and high mass stars respectively and with a turnover at 0.65\,M$_\odot$. The value we have obtained for the slope of the mass function for low-mass stars ($\alpha_{\rm low}=0.85\pm0.10$) is consistent with \citet{Baker} and \citet{Boudreault12}. The presence and location of the turnover in the mass function at $m=0.65$\,M$_\odot$ is in agreement with \citet{Boudreault12} who found that the mass function of Praesepe has a maximum at $~0.6$\,M$_\odot$. We also found that the high-mass slope of the radial mass functions increases from inner to outer radii, providing further evidence for mass segregation in Praesepe. 

From an inspection of the CMD of Praesepe, we also identified 25 binaries and 18 multiple stars. Including unresolved binaries in the CMD, we did a series of simulations to recover the true binary fraction and the true initial mass function. According to our simulations the model which shows the best agreement with the observed mass function and binary fraction has an underlying mass function similar to the observed mass function but an overall binary fraction of $f_{\rm bin}=35\%\pm5\%$. 

\par Finally, we derive a mass of $424$\,M$_\odot$ for Praesepe from the visible stars after subtracting the contribution of contaminants. By considering the contribution of low mass stars, white dwarfs and unresolved binaries this value increases to $\approx 630$\,M$_\odot$.

\section*{Acknowledgments}

H.B. acknowledges support from the Australian Research Council through Future Fellowship grant FT0991052. The authors also would like to thank Wing Cheng and the anonymous referee for their useful comments.

\appendix
\section{Algorithm for finding the mass function slope and break point}\label{sec:powerlaw}
\subsection{Deriving $\alpha$ and its error $\sigma(\alpha)$ }
\noindent For a power law distribution with both lower and upper bounds ($x_{\rm min}$,$x_{\rm max}$) the probability density function is
$$p(x)=Cx^{-\alpha}$$ 
where $C$ is the normalisation constant and $\alpha\neq1$.

\noindent The likelihood of the data given the model with scaling parameter $\alpha$ is
$$p(x|\alpha)=\prod_{i=1}^{n}(1-\alpha)\frac{x_i^{-\alpha}}{\displaystyle x_{\rm max}^{1-\alpha}-\displaystyle x_{\rm min}^{1-\alpha}}$$
\noindent The logarithm $\mathcal L$ of the likelihood is
$${\mathcal L} = \ln p(x|\alpha)=\ln \prod_{i=1}^{n}(1-\alpha)\frac{x_i^{-\alpha}}{\displaystyle x_{\rm max}^{1-\alpha}-\displaystyle x_{\rm min}^{1-\alpha}}$$
The \textit{maximum likelihood estimate} for $\alpha$ is obtained when
$$\frac{\partial\mathcal L}{\partial\alpha}=0$$
\noindent Setting $X=\dfrac{x_{\rm max}}{x_{\rm min}}$, we obtain
\begin{equation*}
\frac{\partial \mathcal L}{\partial \alpha} = \frac{n}{\alpha-1}-\sum_{i=1}^{n}\ln \frac{x_i}{x_{\rm min}}+n\frac{\ln X}{1-X^{\alpha-1}}
\end{equation*}

\begin{equation}\label{Eq:alpha}
\rightarrow \alpha = 1+n\bigg[\sum_{i=1}^{n}\ln\frac{x_i}{x_{\rm min}}-n\frac{\ln X}{1-X^{\alpha-1}}\bigg]^{-1}
\end{equation}

\noindent Eq. \eqref{Eq:alpha} is a general formula which holds for both $\alpha<1$ and $\alpha>1$. 
In general one needs to use numerical methods to calculate $\alpha$ from Eq. \eqref{Eq:alpha} since there is no general analytic solution to this equation.
However, if we assume that there is only a lower bound on data then for $\alpha>1$ the last term of Eq. \eqref{Eq:alpha} vanishes and there is an analytic solution as follows (also given in \citealt{Clauset})
$$\alpha=1+n\bigg[\sum_{i=1}^{n}\ln \frac{x_i}{x_{\rm min}} \bigg]^{-1}$$

\noindent To estimate the error of $\alpha$ we need to calculate the variance of $\alpha$ as explained in \citet{Fisher} which is 
\begin{equation}\label{Eq:var}
\sigma^2(\alpha)=-\bigg(\frac{\partial^2\mathcal L}{\partial \alpha^2}\bigg)^{-1}
\end{equation}

\noindent By substituting $\mathcal L$ into Eq. \eqref{Eq:var} one obtains

\begin{equation}\label{Eq:error}
\sigma(\alpha) = \frac{1}{\sqrt{n}}\Bigg((\alpha-1)^{-2}-\ln^2{X}\frac{X^{\alpha-1}}{(1-X^{\alpha-1})^{2}}\Bigg)^{-1/2} 
\end{equation}

\noindent Again if we assume that there is only a lower bound on data then for $\alpha>1$, Eq. \eqref{Eq:error} reduces to the 
following equation which is given in \citet{Clauset}

$$\rightarrow \sigma(\alpha) = \frac{\lvert\alpha-1\rvert }{\sqrt{n}}$$

\subsection{Finding the mass function turnover (break point)}\label{sec:MFBP}
So far we have derived Eq. \eqref{Eq:alpha} and Eq. \eqref{Eq:error} to estimate $\alpha$ and its error $\sigma(\alpha)$. However, often
mass functions of stellar clusters show a turnover. To find the mass function turnover we proceed as follows:

\begin{itemize}
\item[1.] Decide on the mass range in which the turnover ($m_{\rm t}$) lies. Let $I_{\rm M}$ denote the selected mass range.
\\ \item[2.] Pick a mass (say $m_{\rm x}$) from $I_{\rm M}$.
\\ \item[3.] Fit two power law functions to the data. One from $m_{\rm min}$ to $m_{\rm x}$ and the other 
from $m_{\rm x}$ to $m_{\rm max}${\footnote{$m_{\rm min}$ and $m_{\rm max}$  correspond to the whole mass range and 
not to $I_{\rm M}$ defined in step 1.}}.Calculate the corresponding $\alpha$ values using Eq. \eqref{Eq:alpha}.
\\ \item[4.] Make two theoretical power law distributions using the calculated $\alpha$ values and combine these two 
theoretical distributions to obtain one overall distribution for the whole mass range.
\\ \item[5.] Compare this distribution with the observed distribution using a Kolmogorov-Smirnov test and record the K-S statistic $D$.
\\ \item[6.] Vary $m_{\rm x}$ in $I_{\rm M}$ until $D$ obtains a minimum. The mass for which D becomes minimized 
is the turnover.
\end{itemize}

\par\noindent Following the procedure explained above, one can see that $m_{\rm t}$, $\alpha$ and $\sigma(\alpha)$ are calculated all at once. 

\label{lastpage}

\end{document}